\documentclass[letterpaper, 10 pt, conference]{ieeeconf}
\IEEEoverridecommandlockouts

\usepackage{lipsum}
\usepackage{cite}[sort, compress]
\usepackage[pdftex]{graphicx}
\usepackage{url}
\usepackage{float}
\usepackage{amsmath,amsfonts,amssymb}
\usepackage{xcolor}
\usepackage{mathtools}
\usepackage{theorem}
\usepackage{algorithm}
\usepackage{algpseudocode}
\usepackage{hyperref}
\usepackage{tabularx}
\usepackage{booktabs}

\newcommand{\jac}[1]{\mathrm{J}_{\mathrm{#1}}}

\newcommand{\nx}{n_{\mathrm x}}

\newcommand{\dnu}{n_{\mathrm{u}}}
\newcommand{\np}{n_{\mathrm \eta}}
\newcommand{\D}{\mathcal{D}}
\newcommand{\Real}{\mathbb{R}}

\newcommand{\idx}[2]{\mathbb{I}_{#1}^{#2}}
\DeclareMathOperator{\vectorize}{vec}
\DeclareMathOperator{\mat}{mat}
\newcommand*\crule[3]{%
  {\color[rgb]{#1}\rule[0.4ex]{#2}{#3}}}
\newcommand\dif{\mathop{}\!\mathrm{d}}

{\newtheorem{theorem}{Theorem}}
{\newtheorem{corollary}[theorem]{Corollary}}
{\newtheorem{definition}[theorem]{Definition}}

\title{\LARGE \bf
On-the-fly Surrogation for Complex Nonlinear Dynamics*
}

\author{E. Javier Olucha, Rajiv Singh, Amritam Das and Roland T\'{o}th
\thanks{*This research was supported by the European Space Agency (grant number: 4000145530) and The MathWorks Inc. Opinions, findings, conclusions or recommendations expressed in this paper are those of the authors and do not necessarily reflect the views of The MathWorks Inc. or the European Space Agency.}
\thanks{E.J. Olucha, A. Das and R. T\'{o}th are with the Control Systems Group, Eindhoven University of Technology, The Netherlands. R. Singh is with The MathWorks, Inc., Natick, USA. R. T\'{o}th is also with the Systems and Control Laboratory, HUN-REN Institute for Computer Science and Control, Hungary. Email addresses: \small{\{\tt\small{e.j.olucha.delgado, am.das, r.toth\}@tue.nl}, rsingh@mathworks.com}}
\thanks{Corresponding author: E. Javier Olucha}}%

\begin{document}

\maketitle
\thispagestyle{empty}
\pagestyle{empty}

\begin{abstract}
High-fidelity models are essential for accurately capturing nonlinear system dynamics. However, simulation of these models is often computationally too expensive and, due to their complexity, they are not directly suitable for analysis, control design or real-time applications. Surrogate modelling techniques seek to construct simplified representations of these systems with minimal complexity, but adequate information on the dynamics given a simulation, analysis or synthesis objective at hand. Despite the widespread availability of system linearizations and the growing computational potential of autograd methods, there is no established approach that systematically exploits them to capture the underlying global nonlinear dynamics. This work proposes a novel surrogate modelling approach that can efficiently build a global representation of the dynamics on-the-fly from local system linearizations without ever explicitly computing a model. Using radial basis function interpolation and the second fundamental theorem of calculus, the surrogate model is only computed at its evaluation, enabling rapid computation for simulation and analysis and seamless incorporation of new linearization data. The efficiency and modelling capabilities of the method are demonstrated on simulation examples.
\end{abstract}

\vspace{-5pt}
\section{Introduction\label{sec:intro}}
Over the past century, \emph{linear time-invariant} (LTI) dynamic models have been extensively used in engineering and industrial applications due to their simplicity and effectiveness~\cite{7823045}. Nonetheless, the increasing complexity and performance demands of modern systems often involve operational regimes where nonlinear dynamics inherent to physical processes play a more critical role, exceeding the capabilities of the LTI framework. Consequently, industrial applications require the construction of \emph{high-fidelity} models, which are detailed, precise representations of physical systems, closely replicating their actual behaviour. While these models provide high accuracy, they come at the cost of high model complexity in terms of model order, nonlinear dynamic relationships, dynamic coupling and computational load. This complexity often renders high-fidelity models not directly suitable for simulation, system analysis, controller design or real-time implementations.

These challenges manifested in the need for so-called \emph{surrogate models}, which are low-complexity approximations/representations of all important information about the dynamics of the high-fidelity model for a given utilization objective. Various surrogate modelling frameworks have been developed, such as
\emph{linear parameter-varying} (LPV) systems~\cite{tothModelingIdentificationLinear2010, whiteLinearParameterVaryingControl2013} and \emph{Koopman} approaches~\cite{doi:10.1137/16M1062296, mauroyKoopmanOperatorSystems2020} which aim to provide a simplified linear representation of the nonlinear system with respect to the input-output signal relations. LPV methods based on interpolated linearizations, however, cannot represent transient dynamics between local models.
More generally, these methods trade linearity for conservativeness and scheduling/state order of the resulting surrogate models, often necessitating further model reduction.
Also, conversion methods to LPV and Koopman forms often require either an explicit analytical form of the original model or an extensive trajectory-based representation of the system behaviour.
Another widely used approach involves various types of \emph{neural networks} (NN) or \emph{Gaussian process} (GP) models to learn surrogate models from input-output simulation data~\cite{zhuPhysicsConstrainedDeepLearning2019, 10383457, antonelloPhysicsInformedNeural2023}.
These methods, however, are computationally demanding, their success largely depends on the generation of sufficiently informative data sets, and are inherently approximative in their nature. Moreover, incorporating new data requires retraining the obtained models, which often prevents real-time adaptation of these types of surrogates.

On the other hand, computation of gradients of complicated nonlinear functions has become an inexpensive operation thanks to the recent development of \emph{automatic differentiation} methods~\cite{maclaurin2015autograd}. Consequently, obtaining system snapshots, i.e., local linear approximations of a nonlinear dynamical system at a specific point in the operating space,
is currently a cheap and reliable source of system information.
Each snapshot contains a rich representation of the local system behaviour, and by collecting snapshots across the expected operating range, it is possible to generate an informative data set of the system behaviour. Despite these advantages, figuring out from these local linearizations an accurate global representation of the system behaviour has been found challenging, often considered to be impossible in the literature, and has led to many local approximation methods since the dawn of gain scheduling \cite{zhangLocalLTIModel2020,BACHNAS2014272}.
Hence, there is no established technique that systematically exploits local system snapshots to construct a surrogate model capable of capturing the entire global nonlinear dynamics. Existing approaches are either heavily approximative, rely on complex model conversion methods, or data-driven approximations that demand extensive computational resources and retraining upon the acquisition of new data. This gap motivates the need for a surrogate modelling framework that employs local linear information to reconstruct the global dynamics and allows the incorporation of new system observations without intensive computation processes.

In this work, we introduce a novel light-weight surrogate modelling approach that leverages system snapshots to construct
a surrogate model whose computation is performed ``on-the-fly'', i.e.,
the surrogate model is only computed at its evaluation,
while it is learning the dynamics of the underlying nonlinear system in a global sense. To achieve this, we rely on two key contributions.
First, based on the \emph{second fundamental theorem of calculus} (FTC)~\cite{apostolCalculus1Onevariable1980}, we establish a one-to-one correspondence between the nonlinear dynamics and their local linearization.
Second, we extend the \emph{radial basis functions} (RBF)-based multidimensional \emph{scattered data interpolation} approach~\cite{wendlandScatteredDataApproximation2004,fasshauerMeshfreeApproximationMethods2008} to reconstruct the local linearized dynamics of a system from a finite set of snapshots.
By combining these ingredients with numerical integration, a reliable approximation of the global dynamics at a time moment can be computationally efficiently obtained in the form of a surrogate model from the local dynamics.
Compared to other surrogate modelling methods, our method offers three key advantages: (i) it provides a surrogate model that is not approximative in its nature, and we prove that it converges to the true nonlinear system representation as the number of local linearizations increases; (ii) the construction of the surrogate model
does not require the explicit equations of the underlying system or large data sets but a limited set of local linearizations only,
and (iii) can be updated with new data seamlessly without requiring costly retraining or additional optimization steps.

The paper is structured as follows. In Section~\ref{problem}, the surrogate modelling problem of nonlinear systems based on a set of local linearizations is formulated. For this problem, the proposed on-the-fly approach is presented in Section~\ref{sec:method}, while in Section~\ref{sec:examples}, the capabilities of the method are shown in simulation studies of a controlled Van der Pol oscillator and a mass-spring-damper system with nonlinear dynamics. Finally, in Section~\ref{sec:conclusion}, the main conclusions on the achieved results and further research directions are discussed.

\emph{Notation:} The set of real numbers, integers, and positive integers are denoted as $\Real$, $\mathbb{Z}$, and $\mathbb{Z}^+$, respectively. Let $a,b \in \mathbb{Z}^+$. The row-wise vectorization of a matrix $M\in \Real^{a \times b}$ is given by $\vectorize(M) \in \Real^{1 \times a b}$, while its inverse (under row dimension $a$) is ${\mat_{a}}(\vectorize(M))=M$.
Similarly, consider column-wise vectorization as $\vectorize_{\mathrm{c}}(M)$ and consecutive application of the operation in the form of $\vectorize_{\mathrm{c}}(M_1,\ldots,M_n)$.
The gradient of the function $g:\Real^a \to \Real$ and the Jacobian of $g:\Real^a \to \Real^b$ w.r.t $z$ are denoted by $\nabla_\mathrm{z}(g)$ and $\jac{z}(g)$, respectively.
\vspace{-3pt}
\section{Problem definition\label{problem}}
\subsection{System Description}
Consider a time-invariant nonlinear system $\Sigma$ defined by the state-space representation
\begin{equation} \label{eq:NLmodel}
    \Sigma: \bigl\{ \ \ \xi x(t) = f(x(t), u(t), \eta) ,
\end{equation}
where $t \in \mathbb{T}$ is time, $\xi$ is $\xi x(t) = \frac{d}{dt}x(t)$ in the \emph{continuous-time} (CT) case with $\mathbb{T}=\Real$ and $\xi x(t) = x(t+1)$ in the \emph{discrete-time} case with $\mathbb{T} = \mathbb{Z}$. The signals $x(t) \in \mathbb{X} \subseteq \Real^{\nx}$ and $u(t) \in \mathbb{U} \subseteq \Real^{\dnu}$ with $\nx, \dnu \in \mathbb{Z}^+$ are the states and inputs associated with \eqref{eq:NLmodel}, respectively. Moreover, $\eta \in \mathbb{H} \subseteq \Real^{\np}$ with $\np \in \mathbb{Z}^+$ represents some physical parameters that affect the relation in \eqref{eq:NLmodel}. For the sake of simplicity, we consider the dynamics only in terms of the state evolution in \eqref{eq:NLmodel}; however, one can also consider output equations in the form of $y=h(x(t),u(t))$ with appropriate dimensions.

The function $f: \Real^{\nx} \times \Real^{\dnu} \times \Real^{\np} \to \Real^{\nx}$ is assumed to be at least one-time continuously differentiable, i.e., $f \in \mathcal{C}^k$ with \mbox{$k \in \mathbb{Z}^+$}, and in the CT case, for any initial condition $x(t_0)$ and any input trajectory $u(t)$, the solutions of~\eqref{eq:NLmodel} to be forward complete and unique for all $t\geq t_0$.
Furthermore, let $\mathcal{X} \coloneq \{\underline{x}_i \leq x_i(t) \leq \overline{x}_i \}_{i=1}^{\nx} \subseteq \mathbb{X}$ denote the admissible state space, where $\underline{x}_i$ and $\overline{x}_i$ are the respective lower and upper bounds of the $i$-th state variable, which can be defined based on the operating range of the system. Likewise, let $\mathcal{U} := \{\underline{u}_i \leq u_i(t) \leq \overline{u}_i \}_{i=1}^{\dnu} \subseteq \mathbb{U}$  and $\mathcal{H} := \{\underline{\eta}_i \leq \eta_i \leq \overline{\eta}_i \}_{i=1}^{\np} \subseteq \mathbb{H}$ denote the admissible input and parameter spaces, respectively. {Furthermore, let $z=\vectorize_{\mathrm{c}}(x,u,\eta)$ and also consider the admissible space of $\mathcal{Z}=\mathcal{X}\times\mathcal{U}\times\mathcal{H}$.}
\vspace{-3pt}
\subsection{Data collection}
Let $\mathcal{L}_f : \Real^{\nx} \times \Real^{\dnu} \times \Real^{\np} \to \Real^{\nx \times (\nx + \dnu)}$ be the linearization operator of $f$, i.e., the operator that provides the linearized dynamics of~\eqref{eq:NLmodel} at a point $(x_\ast, u_\ast, \eta_\ast) \! \in \! \mathcal{X} \! \times \mathcal{U} \!\times \mathcal{H}$:
\vspace{-3pt}
\begin{equation} \label{eq:linearization}\vspace{-3pt}
    \mathcal{L}_f(x_\ast, u_\ast, \eta_\ast)  \coloneq \begin{bmatrix}
        \jac{x} (f) & \jac{u} (f)
    \end{bmatrix}(x_\ast, u_\ast, \eta_\ast).
\end{equation}
Consider a set of local snapshots, i.e., linearizations of \eqref{eq:NLmodel}:\vspace{-2pt}
\begin{equation} \label{eq:data}\vspace{-2pt}
    \mathcal{M} \coloneq  \{\underbrace{\mathcal{L}_f(x^{(i)}, u^{(i)}, \eta^{(i)})}_{M_i}\}_{i=1}^N,
\end{equation}
where $\Theta  \coloneq \{z^{(i)}\}_{i=1}^N $ with $z^{(i)}=\vectorize_{\mathrm{c}}(x^{(i)}, u^{(i)}, \eta^{(i)})$ is a set of $N \in \mathbb{Z}^+$ distinct observation points. 
The collection of available local information about \eqref{eq:NLmodel} is denoted by
\begin{equation}\label{eq:dataset}\vspace{-3pt}
    \D \coloneq \{\Theta, \mathcal{M}\}.
\end{equation}
\subsection{Objective}
Based on these considerations, our objective is to find a surrogate representation of the system in the form of
\begin{equation}\label{eq:surrogate}\vspace{-3pt}
    \hat\Sigma: \bigl\{ \ \ \xi \hat{x}(t) = \hat{f}(\hat{x}(t), u(t), \eta),
\end{equation}
where at any time moment $t$, i.e., for any value of $(\hat{x}(t), u(t), \eta)$, the true nonlinear function $f(\hat x(t), u(t), \eta)$ corresponding to the state increment $\xi \hat{x}(t)$ is approximated with the value $\hat{f}$ that is directly calculated based on the set of local linearizations $\D$ only.

Note that no analytic form of the mapping $\hat{f}:\Real^{\nx} \times \Real^{\dnu} \times \Real^{\np} \to \Real^{\nx}$ is required, only the true relation $f(\hat x(t), u(t), \eta)$ is aimed to be represented with (i) minimal approximation error and (ii) computationally efficient calculation of $\hat{f}(\hat x(t), u(t), \eta)$ based on $\D$.
\section{Methodology\label{sec:method}}
\subsection{From local linearizations to global representation\label{sec:ftc}}
As a first step towards our objective, we establish a key relationship between the nonlinear state-space representation and its linearized dynamics. Building on the ideas in~\cite{koelewijnAnalysisControlNonlinear2023},~\cite{oluchaAutomatedLinearParameterVarying2025}, and utilizing the FTC, we show that the true nonlinear state equation given in~\eqref{eq:NLmodel} can be recovered completely from its linearizations given by $\mathcal{L}_f$ in~\eqref{eq:linearization}.
\vspace{-5pt}
\begin{theorem}\label{thm:nonlinear_linear}
    Given the nonlinear dynamics~\eqref{eq:NLmodel}, the following identity holds true for any $(x_\ast, u_\ast) \in \mathbb{X} \times \mathbb{U}$: 
    \begin{equation}
        \label{eq:nl_to_linear}\vspace{-2pt}
        f(x, u, \eta) = f(x_\ast, u_\ast, \eta) + F(x, u, \eta) \begin{bmatrix}
            x - x_\ast \\ u - u_\ast
        \end{bmatrix},
    \end{equation}
    where
    \vspace{-3pt}\begin{align*}
        F(x, u, \eta) \hspace{-1pt} & = \hspace{-3.5pt} \int\limits_{0}^{1} \hspace{-3.5pt} \mathcal{L}_f \left(x_\ast + \lambda (x - x_\ast), u_\ast + \lambda (u - u_\ast), \eta\right) \, \dif \lambda.
    \end{align*}
\end{theorem}
\begin{proof}
    For any nonlinear function $g_i: \Real^{n} \rightarrow \Real$, $g_i\in\mathcal{C}^{1}$, scalar $\lambda \in [0,1]$ and point $z_{*} \in \Real^{n}$,
    we define the following auxiliary function $\bar{g}$ as follows
    \begin{equation}
        \label{eq:auxiliary}
        \bar{g}_i(z;\lambda) = g_i(z_{\ast} + \lambda(z - z_{\ast})).
    \end{equation}
    By the FTC and the identities $\bar{g}_i(z;1)=g_i(z)$, $\bar{g}_i(z;0) =  g_i(z_{\ast})$:
    \begin{equation}
        \label{eq:FTC2}
        g_i(z) - g_i(z_{\ast}) = \int\limits_{0}^{1} \frac{\dif \bar{g}_i}{\dif \lambda}(z;\lambda) \, \dif \lambda.
    \end{equation}
    Now, applying the chain rule of the integrand in \eqref{eq:FTC2}:
    \begin{equation}
        \label{eq:chain_rule}
        \frac{\dif \bar{g}_i}{\dif \lambda}(z;\lambda) = \Bigg[\Big(\nabla_\mathrm{z} g_i\Big)^{\top}\Big(z_{*} + \lambda(z-z_{*})\Big)\Bigg] (z - z_{\ast}).
    \end{equation}
    Substitution of \eqref{eq:chain_rule} into \eqref{eq:FTC2} yields
    \begin{equation}
        \label{ftc_element}
        g_i(z) = g_i(z_{\ast}) + \int\limits_{0}^{1} \Bigg[\Big(\nabla_\mathrm{z} g_i\Big)^{\top}\Big(z_{*} + \lambda(z-z_{*})\Big)\Bigg] \, \dif \lambda \, (z - z_{\ast}).
    \end{equation}
    Then, consider $g(z) := [g_1(z) \ \cdots \ g_m(z)]^\top$. By stacking \eqref{ftc_element} column-wise for all $i\in [1, \cdots, m]$, $g$ is expressed as
    \begin{equation}
        \label{eq:fun_factorization}
        g(z) = g(z_{\ast}) + \int_{0}^{1} \jac{z}\Big(g(z_{\ast} + \lambda(z - z_{\ast}))\Big) \, \dif \lambda \, (z - z_{\ast}).
    \end{equation}
    Now, by applying~\eqref{eq:fun_factorization} to the function $f$ in~\eqref{eq:NLmodel} leads to~\eqref{eq:nl_to_linear}.
\end{proof}

When $f(0, 0, \eta) = 0$, which corresponds to an equilibrium point in CT, Theorem~\ref{thm:nonlinear_linear} can be simplified.
\vspace{-5pt}
\begin{corollary}\label{thm:nonlinear_linear_simplified}
    If $f(0, 0, \eta) = 0$ for a fixed $\eta \in \mathbb{P}$, then
    the right-hand side of~\eqref{eq:NLmodel} admits the following expression in terms of the linear dynamics:
    \begin{equation}    \label{eq:simplified_factorized_nl}\vspace{-3pt}
        f(x, u, \eta) = F(x, u, \eta) \begin{bmatrix}
            x \\ u 
        \end{bmatrix},
    \end{equation}
    where
    \vspace{-3pt}\begin{align*}
        F(x, u, \eta) & = \int_{0}^{1} \mathcal{L}_f \left(\lambda x, \lambda u, \eta\right) \, \dif \lambda.
    \end{align*}
\end{corollary}
\begin{proof}
    The proof is straightforward by substituting $(x_\ast, u_\ast, \eta) = (0, 0, \eta)$ into~\eqref{eq:nl_to_linear} and using $f(0, 0, \eta) = 0$.
\end{proof}

The results from Corollary~\ref{thm:nonlinear_linear_simplified} allow the exact realization of the true nonlinear function $f$ from the linearization operator $\mathcal{L}_f$, provided $f(0, 0, \eta) = 0$. Note that when
\mbox{$f(0, 0, \eta) \neq 0$}, but $(0, 0) \in \mathcal{X} \times \mathcal{U}$, we can perform a coordinate transformation to ensure this property. 
However, application of Corollary~\ref{thm:nonlinear_linear_simplified} has a major drawback: it requires complete knowledge of the linearization of $f$, i.e., the operator $\mathcal{L}_f$, over the entire operating space. As in our problem setting we only have a finite number of local snapshots of the system, we need to approximate $\mathcal{L}_f$ from the set of linearizations $\D$. This is what we tackle in the next subsection.

\vspace{-2pt}\subsection{Interpolation of local system dynamics\label{sec:rbf}}
The system linearizations in $\D$~\eqref{eq:dataset} provide exact representations of the original system \eqref{eq:NLmodel} at their respective linearization points. The surrogation problem then amounts to approximating $\mathcal{L}_f$ from $\D$ while preserving this exactness, which naturally leads to the following interpolation problem.
\subsubsection{Formulation of the interpolation problem}
For a given set $\D$, find an interpolant $\mathcal{I}^\mathcal{\D} : \Real^{\nx} \times \Real^{\dnu} \times \Real^{\np} \to \Real^{\nx \times (\nx + \dnu)}$ such that the residual $r: \Real^{\nx} \times \Real^{\dnu} \times \Real^{\np} \to \Real$ between  $\mathcal{L}_f$ and $\mathcal{I}^\mathcal{\D}$, defined by
\begin{equation}\label{eq:residual}
    r(z) \coloneq \| \mathcal{L}_f(z) - \mathcal{I}^\mathcal{\D}(z) \|_2,
\end{equation}
satisfies the following conditions:
\begin{subequations}\label{eq:residual_conditions}
    \begin{align}\label{eq:interpolation_condition}
        r(z) & = 0  \quad \text{for all} \quad z^{(i)} \in \Theta,                                                              \\
        r(z) & < \epsilon \quad \text{for all} \quad z^{(i)} \in \mathcal{Z} \setminus \Theta. \label{eq:intersample_condition}
    \end{align}
\end{subequations}

The problem of determining $\mathcal{I}^\mathcal{\D}$ subject to~\eqref{eq:residual_conditions} is a multidimensional interpolation problem suggesting that most regression-based methods,
such as~\cite{Tibshirani:1996fxl, specht1991general},
are not suitable since they do not satisfy~\eqref{eq:interpolation_condition}.
Furthermore, one must not assume that the observation points contained in $\D$ are uniformly distributed, as they can be scattered in reality.

\subsubsection{Multidimensional scattered interpolation with RBFs} Although multiple approaches exist to construct $\mathcal{I}^\mathcal{\D}$ from the dataset $\D$, we propose the use of RBFs, as they provide an efficient and robust solution for multidimensional scattered interpolation~\cite{wu_local_1993,amidrorScatteredDataInterpolation2002}.
The RBF-based interpolant that approximates $\mathcal{L}_f$ is given by
\vspace{-3pt}\begin{equation}\label{eq:RBFinterpolant}
    \mathcal{I}_{\mathrm{RBF}}^\mathcal{\D}(z) = \mat_{\nx}(\sum_{i=1}^{N} \alpha_i \phi(z - z^{(i)})),
\end{equation}
where $\phi$ is a RBF, i.e., a continuous function $\phi : \Real^d \to \mathbb{C}$ with $d = \nx + \dnu + \np$ which is invariant under translation and rotation, $\alpha \in \Real^{N \times \nx (\nx + \dnu)}$, and the interpolation condition in~\eqref{eq:interpolation_condition} is imposed as
\begin{equation}\label{eq:interpolation_condition_RBF}
    \mathcal{I}_{\mathrm{RBF}}^\mathcal{\D}(z^{(i)}) = M_i \quad \text{for} \quad i \in \idx{1}{N},
\end{equation}
where $\idx{a}{b} \coloneq \{i\in\mathbb{Z}^+ \mid a \leq i \leq b\}$ is an index set. Then, the coefficients $\alpha_i$ can be determined by formulating~\eqref{eq:interpolation_condition_RBF} as
\begin{subequations}\label{eq:RBF_system}
    \begin{equation}
        R \; \alpha = \gamma, \vspace{-5pt}
    \end{equation}
    where
    \begin{align}
         & [R]_{i, j}  \coloneq \phi(z^{(i)} - z^{(j)}), \quad i, j \in \idx{1}{N}, \quad R\in \Real^{N \times N},                               \label{eq:RBF_R} \\
         & [\gamma]_i  \coloneq \vectorize(M_i), \quad i \in \idx{1}{N}, \quad \gamma \in \Real^{N \times {\nx (\nx + \dnu)}}. \label{eq:RBF_gamma}
    \end{align}
\end{subequations}

\subsubsection{Extension of the RBF interpolant with a polynomial tail} When variations of the local system dynamics are described by polynomial relationships of total degree less than $m$, the RBF-based interpolant~\eqref{eq:RBFinterpolant} can be improved by adding a polynomial tail~\cite{wendlandScatteredDataApproximation2004}. This polynomial tail represents the basis of $d$-variate polynomials of total degree at most $m-1$ which guarantees polynomial precision.
The polynomial tail, denoted by $\mathcal{I}_{\mathrm{poly}}^{\D}$, is defined as
\vspace{-2pt}\begin{equation}\label{eq:RBFpoly}\vspace{-2pt}
    \mathcal{I}_{\mathrm{poly}}^{\D}(z) = \mat_{\nx}(\sum_{j=1}^{Q} \beta_j q_j(z)),
\end{equation}
where $\beta \in  \Real^{Q \times \nx (\nx + \dnu)}$ and $q_1, \dots, q_Q$ is a basis for the $d$-variate polynomials of degree $\leq m-1$ with dimension $Q = {(m-1+d)!} / {((m-1)! \; d!)}$.
In this case, the interpolant that approximates $\mathcal{L}_f$ becomes
\begin{equation}\label{eq:CPD_interpolant}\vspace{-2pt}
    \mathcal{I}^{\D}(z) = \mathcal{I}_{\mathrm{RBF}}^\mathcal{\D}(z) + \mathcal{I}_{\mathrm{poly}}^{\D}(z),
\end{equation}
where now the interpolation condition~\eqref{eq:interpolation_condition}
\begin{equation} \label{eq:CPD_interpolation_condition}\vspace{-2pt}
    \mathcal{I}^{\D}(z^{(i)})                         = M_i, \quad i \in \idx{1}{N}, \\
\end{equation}
is complemented with the side condition
\begin{equation}\label{eq:CPD_side_conditions}
    \sum_{i}^{N} \alpha_i q_k(z^{(i)}) = 0, \quad k \in \idx{1}{Q},
\end{equation}
to cope with the additional degrees of freedom introduced by the polynomial tail.

\subsubsection{Determining coefficients of the interpolant}
The coefficients $\alpha_i$ and $\beta_j$ related to $\mathcal{I}_{\mathrm{RBF}}^\mathcal{\D}$ and $\mathcal{I}_{\mathrm{poly}}^{\D}$ can now be determined by extending~\eqref{eq:RBF_system} with the side conditions~\eqref{eq:CPD_side_conditions}, leading to:
\begin{subequations}\label{eq:CPD_system}
    \begin{equation}
        \label{eq:CPD_new}\vspace{-2pt}
        \begin{bmatrix}
            R & P \\ P^\top & 0
        \end{bmatrix} \begin{bmatrix}
            \alpha \\ \beta
        \end{bmatrix} = \begin{bmatrix}
            \gamma \\ 0
        \end{bmatrix},
    \end{equation}
    with
    \begin{align} \label{eq:CPD_P}
        [P]_{i, j} & \coloneq q_j(z^{(i)}), \quad i\in \idx{1}{N}, \ j \in \idx{1}{Q}, \ P \in \Real^{N \times Q},
    \end{align}
\end{subequations}
where $R$ and $\gamma$ are defined as in~\eqref{eq:RBF_system}.
Now, we can show the uniqueness of solutions of~\eqref{eq:CPD_system} on the basis of~\cite[Chapter 8]{wendlandScatteredDataApproximation2004}. To this end, we first introduce the notion of \emph{conditionally positive definite} (CPD) functions~\cite[Definition 8.1]{wendlandScatteredDataApproximation2004}.
\vspace{-5pt}\begin{definition} \label{thm:CPDfunctions}
    A continuous, even function $\phi:\Real^d \to \Real$ is said to be CPD of order $m$ if, for all pairwise distinct observation points $z_1, \dots, z_N \in \Real^d$ with $N \in \mathbb{Z}^+$ and all $\alpha \in \Real^N \times \Real^{\nx \times (\nx + \dnu)} \setminus \{0\}$ that satisfy
    \begin{equation}
        \sum_{i=1}^{N} \alpha_i q(z^{(i)}) = 0
    \end{equation}
    for all real-valued polynomials q of degree less than $m$, the following inequality holds that
    \begin{equation}
        \sum_{i=1}^{N} \sum_{j=1}^{N} \alpha_i \alpha_j \phi(z^{(i)} - z^{(j)}) > 0.
    \end{equation}
\end{definition}
Then, using Definition~\ref{thm:CPDfunctions}, we have the following result that closely follows the treatment of~\cite[Theorem 8.21]{wendlandScatteredDataApproximation2004}.
\vspace{-5pt}\begin{theorem}
    \label{thm:CPD_uniqueness}
    Suppose that the function $\phi$ is CPD of order $m$ and the columns of $P$ are linearly independent, where $P$ is given in~\eqref{eq:CPD_P}. Then,~\eqref{eq:CPD_system} has a unique solution.
\end{theorem}
\begin{proof}
    Suppose that $[\alpha^\top \ \beta^\top]^\top \in \Real^{(N+Q) \times \nx (\nx + \dnu)}$ is a solution of the homogeneous equation system in~\eqref{eq:CPD_system}, i.e., with $\gamma = 0$. Then, we have
    \vspace{-3pt}\begin{equation}\label{eq:homogeneous_sys}\vspace{-3pt}
        \begin{aligned}
            R \alpha + P \beta & = 0, \\
            P^\top \alpha      & = 0,
        \end{aligned}
    \end{equation}
    where the second equation means that $\alpha$ satisfies the side condition~\eqref{eq:CPD_side_conditions}. Now, pre-multiplying the first equation by $\alpha^\top$, giving $\alpha^\top \! R \alpha + \alpha^\top \!P \beta = \alpha^\top \! R \alpha +  \left(P^\top \! \alpha\right)^{\!\top} \! \! \beta = \alpha^\top \! R \alpha = 0$. As $\phi$ is CPD of order $m$, we can conclude that $\alpha = 0$ and thus $P \beta = 0$. Then, since the columns of $P$ are linearly independent, we can conclude that $\beta = 0$. Therefore,~\eqref{eq:homogeneous_sys} has trivial solutions only, meaning that the left-hand-side matrix of~\eqref{eq:CPD_new} is full rank, and thus~\eqref{eq:CPD_system} has a unique solution.
\end{proof}

By Theorem~\ref{thm:CPD_uniqueness}, if the RBF is CPD and the data points $z^{(i)}$ yield a full-rank P in~\eqref{eq:CPD_P}, then~\eqref{eq:CPD_system} has a unique solution, ensuring that for a given $\mathcal{D}$ there exists a unique interpolant of the form~\eqref{eq:CPD_interpolant} that satisfies~\eqref{eq:interpolation_condition}.
One example of a CPD RBF function is the \emph{Hardy multiquadric}~\cite{wendlandScatteredDataApproximation2004}, which is CPD of (minimum) order $m = 1$, given by
\begin{equation}\label{eq:RBFhardy}
    \phi(z; c) = -\left(c^2 + \|z\|_2^2 \right)^{1/2},
\end{equation}
where $c>0$ is a constant shape parameter.

Now, for a given data set $\D$, we can use~\eqref{eq:RBFinterpolant} to approximate the linearization operator $\mathcal{L}_f$ within the admissible space. Additionally, if the nonlinear system~\eqref{eq:NLmodel} is expected to contain polynomial relationships in its dynamics, we can use~\eqref{eq:CPD_interpolant} instead to guarantee polynomial precision.
In either case, the construction of the interpolant $\mathcal{I}^{\D}$ only involves an unconstrained linear system of equations, which can be efficiently solved with standard linear least squares methods~\cite{BJORCK1990465}.
In the next section, we detail how to utilize the proposed interpolant $\mathcal{I}^{\D}$ in combination with the results from Theorem~\ref{thm:nonlinear_linear} to achieve our objective of constructing the surrogate model on-the-fly.

\subsection{Finding the surrogate representation on-the-fly}
Given the set $\D$, we can find a light-weight surrogate representation of~\eqref{eq:NLmodel} which is assembled on-the-fly. The first step of the process requires the pre-computation of the interpolant $\mathcal{I}^{\D}$ given by~\eqref{eq:CPD_interpolant}, which is executed in an offline fashion, as detailed in Algorithm~\ref{alg:interp}. Second, assuming that $f(0, 0, \eta) = 0$, we use the results in Corollary~\ref{thm:nonlinear_linear_simplified} with $\mathcal{I}^{\D}$ to formulate the surrogate representation $\hat{\Sigma}$ as follows:
\begin{subequations}
    \begin{equation}\label{eq:surrogateRep}\vspace{-5pt}
        \hat{\Sigma} \coloneq \bigl\{\xi \hat{x} = \hat{f}(\hat{x}, u, \eta) = \hat{F}(\hat{x}, u, \eta) \begin{bmatrix}
            \hat{x} \\ u
        \end{bmatrix},
    \end{equation}
    where\vspace{-7pt}
    \begin{align}\label{eq:interpolated_integral}
        \hat{F}(\hat{x}, u, \eta) \hspace{-1pt} & = \hspace{-3.5pt} \int_{0}^{1} \hspace{-3.5pt} \mathcal{I}^{\D} \left(\lambda \hat{x}, \lambda u, \, \eta\right) \, \dif \lambda,
    \end{align}
\end{subequations}
and the integral in~\eqref{eq:interpolated_integral} is left unevaluated. This representation allows assembling the surrogate model on-the-fly, i.e., the surrogate model is only computed when needed by solving the definite integral in~\eqref{eq:interpolated_integral}, as summarized in Algorithm~\ref{alg:otfs}.
This is a key advantage for real-world applications, since incorporating new system observations only requires the recomputation of $\mathcal{I}^{\D}$.
To this end, it is necessary to invoke a numerical integration scheme to solve~\eqref{eq:interpolated_integral} in a computationally efficient manner, as an analytic solution can be costly and often not computable.
\begin{algorithm}[t]
    \caption{Pre-computation of the interpolant (offline)}\label{alg:interp}
    \textbf{Input:} Data set $\D$ of the form~\eqref{eq:dataset}, RBF function $\phi(z; c)$, width parameter $c$, polynomial basis $q(z)$ of total degree $m$. \\
    \textbf{Output:} Interpolant $\mathcal{I}^{\D}$ over $\D$.
    \begin{algorithmic}[1]
        \State Compute the matrices $R$, $P$ and $\gamma$, given by~\eqref{eq:RBF_R},~\eqref{eq:CPD_P}, and~\eqref{eq:RBF_gamma}, respectively.
        \State Find $\alpha$ and $\beta$ by solving~\eqref{eq:CPD_system}.
        \State Construct the interpolant $\mathcal{I}^{\D}$ given by~\eqref{eq:CPD_interpolant}.
        \State \textbf{Return} $\mathcal{I}^{\D}$.
    \end{algorithmic}
\end{algorithm}
\begin{algorithm}[t]
    \caption{On-the-fly surrogation (online)}\label{alg:otfs}
    \textbf{Input:} Interpolant $\mathcal{I}^{\mathcal{\D}}$, value of $(\hat{x}(t), u(t), \eta)$ at a time moment $t$, numerical integration scheme. \\
    \textbf{Output:} Approximated value of the state increment $\xi \hat{x}(t)$.
    \begin{algorithmic}[1]
        \State Compute $\hat{F}(\hat{x}(t), u(t), \eta)$ in~\eqref{eq:interpolated_integral} with the numerical integration scheme.
        \State Compute $\xi \hat{x}(t)$ by~\eqref{eq:surrogateRep}.
        \State \textbf{Return} $\xi \hat{x}(t)$.
    \end{algorithmic}
\end{algorithm}
In addition, the proposed approach guarantees that $\hat{f}$ converges to the true nonlinear function $f$ as the number of observations in $\D$ increases, leading to the following result.
\vspace{-8pt}\begin{theorem}\label{thm:universalApprox}
    Given a nonlinear system $\Sigma$ defined by~\eqref{eq:NLmodel} where $f\in \mathcal{C}^k$ with $k\in\mathcal{Z}^+$, then for a set $\D_N = \{\Theta, \mathcal{M}\}$ given by~\eqref{eq:dataset} with $N\in\mathbb{Z}^+$ distinct linearizations and the interpolant $\mathcal{I}^{\D}$ given by~\eqref{eq:CPD_interpolant} that approximates $\mathcal{L}_f$ according to~\eqref{eq:linearization}, the surrogate $\hat{f}$ given by~\eqref{eq:surrogateRep}, satisfies
    \begin{equation}
        \|f(z_\ast) - \hat{f}(z_\ast)\| \to 0 \quad
    \end{equation}
    at any point $(z_\ast) \in \mathcal{Z}$ as $N \to \infty$.
\end{theorem}\vspace{-5pt}
\begin{proof}
    The proof follows from the fact that $\mathcal{I}^{\D}$ satisfies the interpolation condition in~\eqref{eq:interpolation_condition}, therefore $\| \mathcal{L}_f - \mathcal{I}^{\D}\| < \epsilon$ at any point $z \in \mathcal{Z}_{\Theta}^\mathrm{c}$, where $\mathcal{Z}_{\Theta}^\mathrm{c}$ denotes the relative complement of $\mathcal{Z}$ w.r.t. $\Theta$, i.e., \mbox{$\mathcal{Z}_{\Theta}^\mathrm{c} \coloneq \mathcal{Z} \setminus \Theta$}. Then, as $N \to \infty$, $\#\mathcal{Z}_{\Theta}^\mathrm{c} \to 0$, where $\#$ denotes the cardinality of the set. Thus, $\| \mathcal{L}_f - \mathcal{I}^{\D}\| \to 0$, and consequently, $\|f(z_\ast) - \hat{f}(z_\ast)\| \to 0$ by Theorem~\ref{thm:nonlinear_linear}.
\end{proof}
\vspace{-3pt}\subsection{Optimization of the RBFs width parameter\label{sec:hyperparameter_tuning}}
The RBF-based interpolation depends on the width $c$ (see~\eqref{eq:RBFhardy}), which influences the approximation quality of the interpolant $\mathcal{I}^{\D}$ and, consequently, the accuracy of the surrogate $\hat{f}$. To tune $c$, we use the ``leave-one-out'' cross-validation based optimization from~\cite{fasshauerMeshfreeApproximationMethods2008}.
For a given $\D_N = \{z^{(k)}, M_k\}_{k=1}^N$, an RBF $\phi(z; c)$ and a polynomial basis $q$ of total degree $m$,
\begin{equation}\label{eq:optimization}\vspace{-5pt}
    c^\ast = \arg \min_{c} \| [E_1 \ldots E_N](c) \|_p,
\end{equation}
with \vspace{-5pt}
\begin{gather*}
    E_k(c)                = M_k - I^{\mathcal{D}_k}(z^{(k)}; c), \quad k \in \idx{1}{N},
\end{gather*}
where $I^{\mathcal{D}_k}$ denotes the interpolant given by~\eqref{eq:CPD_interpolant} over the set $\mathcal{D}_k \coloneq \D \setminus \{z^{(k)}\}$, and $\| \cdot \|_p$ denotes the induced $p$ matrix norm. In particular, $p=\infty$ corresponds to~\eqref{eq:intersample_condition}. Note that~\eqref{eq:optimization} is nonlinear due to the norm and the matrix inversion required for $I^{\mathcal{D}_k}$, and non-smooth when $p = \infty$.
\section{Examples\label{sec:examples}}
In this section, we present two numerical examples to show the capabilities of the proposed on-the-fly surrogation approach. The first example considers the Van der Pol oscillator to provide a clear and intuitive demonstration of the method. The second example involves an interconnection of mass-spring-damper (MSD) systems to demonstrate scalability.
The full implementation is available at: \href{https://gitlab.com/Javi-Olucha/cdc25-code-repo}{https://gitlab.com/Javi-Olucha/cdc25-code-repo}.
\subsection{Van der Pol oscillator}
In this example, we consider the Van der Pol oscillator, whose dynamics are described as follows:
\begin{equation}\label{eq:van_der_pol}
    \Sigma_{\mathrm{vdp}} \coloneq \left\{\begin{bmatrix}
        \dot{x}_1 \\ \dot{x}_2
    \end{bmatrix} = \begin{bmatrix}
        x_2 \\ - x_1 - \eta x_2 (1-x_1^2)
    \end{bmatrix} + \begin{bmatrix}
        0 \\ x_1 u
    \end{bmatrix} \right.,
\end{equation}
where $\eta > 0$ is a system parameter. Moreover, let $\Sigma_{\mathrm{vdp}}^{{\eta}}$ denote the Van der Pol oscillator with the parameter $\eta$ fixed at ${\eta} > 0$. Furthermore, we consider the infinite-horizon optimal control law given by $u = -x_1 x_2$ for the performance objective \mbox{$J(x, u) = x_2^2 + u^2$} proposed by~\cite{nevistić_primbs_1996} that drives the system to the origin.
\subsubsection{Reconstruction of the true nonlinear system from the linear dynamics} \label{sec:VDPexperiment1}
\begin{figure}[t]
    \vspace{-5pt}
    \includegraphics[width=1\columnwidth]{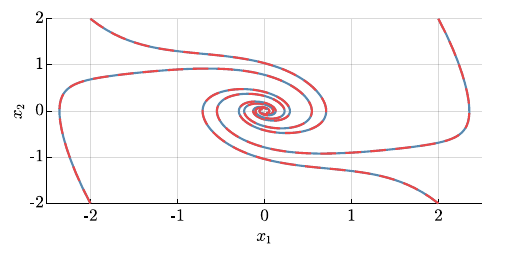}
    \vspace{-26pt}
    \caption{Numerical illustration of Theorem~\ref{thm:nonlinear_linear} using complete knowledge of the linearization operator. The state responses of the CT simulation of $\Sigma_{\mathrm{vdp}}^{0.5}$ are shown, where (\crule{0.3467, 0.5360, 0.6907}{6pt}{1pt}) is obtained with the true system representation and (\crule{0.9153, 0.2816, 0.2878}{3pt}{1pt} \crule{0.9153, 0.2816, 0.2878}{3pt}{1pt}) with its reconstruction using Corollary~\ref{thm:nonlinear_linear_simplified}.
    }\label{fig:NLreconstruction} \vspace{-10pt}
\end{figure}

First, we numerically demonstrate Theorem~\ref{thm:nonlinear_linear}
with the reconstruction of the global dynamics of the Van der Pol oscillator $\Sigma_{\mathrm{vdp}}^{0.5}$ based on perfect reconstruction of the underlying local dynamics\footnote{For complex models, obtaining the local linearized dynamics at any point in the admissible space is too expensive in practise, and only a limited set of local linearizations is available.}.
In this case, as the origin of $\Sigma_{\mathrm{vdp}}^{0.5}$ is an equilibrium, we use Corollary~\ref{thm:nonlinear_linear_simplified}. Then,
we execute a CT simulation of $\Sigma_{\mathrm{vdp}}^{0.5}$ for each of the initial conditions $(-2, -2)$, $(-2, 2)$, $(2, -2)$ and $(2, 2)$.
For the simulations, we use the \textsc{Matlab} built-in variable-step solver \texttt{ode45} with the default parameters and step size, and simulate for 14 seconds. Next, we replicate the simulations using the same settings, but this time we use~\eqref{eq:simplified_factorized_nl} instead of the true nonlinear system $\Sigma_{\mathrm{vdp}}^{0.5}$. The integral in~\eqref{eq:simplified_factorized_nl} is solved numerically using the \emph{composite Simpson's 3/8 rule} (CS38)~\cite[Chapter 5.1]{atkinsonIntroductionNumericalAnalysis1989} with 6 equidistant intervals. For these simulations, \texttt{ode45} requested an average of 852 evaluations of the Jacobian of~\eqref{eq:van_der_pol} with an average computation time of 0.0035 seconds\footnote{On a laptop with an i7-13850HX (2.10 GHz) CPU and 64 GB RAM.}. The simulation results, shown in Fig.~\ref{fig:NLreconstruction}, illustrate that the solutions of $\Sigma_{\mathrm{vdp}}^{0.5}$ and its reconstruction with Corollary~\ref{thm:nonlinear_linear_simplified} are identical, with an average \emph{root-mean-square error} (RMSE) of $2.38\times 10^{-16}$, supporting the results in Subsection~\ref{sec:ftc}.
\subsubsection{On-the-fly surrogation of the Van der Pol oscillator}\label{sec:VDPexperiment2}
We next learn surrogate models of $\Sigma_{\mathrm{vdp}}^{0.5}$ using the proposed on-the-fly surrogation method. For this, we use the sets $\Theta^{1}_{\mathrm{vdp}} \subset \Theta^{2}_{\mathrm{vdp}} \subset \Theta^{3}_{\mathrm{vdp}}$ defined in Tab.~\ref{tab:vdp_observation_sets}, which contain
9, 15 and 25
observation points, respectively.
\begin{table}[b]
    \centering
    \caption{Definition of different sets with observation points used to obtain different surrogate models of $\Sigma_{\mathrm{vdp}}^{0.5}$}\label{tab:vdp_observation_sets}\vspace{-7pt}
    \begin{tabular}{l}
        \midrule
        $\Theta^{1}_{\mathrm{vdp}} = \{(x_1,x_2) \mid x_1 \in \{-2,0,2\}, \, x_2 \in \{-2,0,2\}\}$                    \\
        $\Theta^{2}_{\mathrm{vdp}} = \{(x_1,x_2) \mid x_1 \in \{-2,0,2\}, \, x_2\in\mathbb{Z},\; -2\le x_2\le 2 \}$   \\
        $\Theta^{3}_{\mathrm{vdp}} = \{(x_1,x_2) \mid x_1,x_2\in\mathbb{Z},\; -2\le x_1\le 2,\; -2\le x_2\le 2 \} \}$ \\
        \bottomrule
    \end{tabular}
\end{table}
From these, we construct the dictionaries $\D_{\mathrm{vdp}}^i = \{\Theta_{\mathrm{vdp}}^i, \mathcal{M}_{\mathrm{vdp}}^i\}$ with $i=1,2,3$, where the sets $\mathcal{M}_{\mathrm{vdp}}^{i}$ are obtained by linearizing $\Sigma_{\mathrm{vdp}}^{0.5}$ at the points in $\Theta_{\mathrm{vdp}}^{i}$.
The interpolants $\mathcal{I}_{\mathrm{vdp}}^{\D_1}$, $\mathcal{I}_{\mathrm{vdp}}^{\D_2}$ and $\mathcal{I}_{\mathrm{vdp}}^{\D_3}$ are then computed using Algorithm~\ref{alg:interp} with the RBF $\phi$ given in~\eqref{eq:RBFhardy} and the polynomial basis $\mathcal{I}_{\mathrm{poly}}(x, u) = \beta_1 + \beta_2 x_1 + \beta_3 x_2$. The width parameter $c$ is determined by minimizing~\eqref{eq:optimization} with {Bayesian optimization}~\cite{brochuTutorialBayesianOptimization2010}, given that~\eqref{eq:optimization} is non-smooth. Other non-smooth optimization schemes could also be employed. Next, the OTFS surrogate representations $\hat{\Sigma}_{\mathrm{vdp}}^{\D^1}$, $\hat{\Sigma}_{\mathrm{vdp}}^{\D^2}$ and $\hat{\Sigma}_{\mathrm{vdp}}^{\D^3}$ are built with Algorithm~\ref{alg:otfs}, the CS38 numerical integration rule with 6 intervals and the corresponding interpolants.

For comparison, we compute the sixth- and tenth-order Koopman surrogates $\hat{\Sigma}_{\mathrm{vdp}}^{K_6}$ and $\hat{\Sigma}_{\mathrm{vdp}}^{K_{10}}$, using the data-driven method of~\cite{williamsDataDrivenApproximation2015}. Both models are trained with the \textsc{PyKoopman} library~\cite{Pan2024} on the trajectories of Fig.~\ref{fig:NLreconstruction} sampled with a time interval of $0.01$ seconds. The observables $g(x) = [1, x_1, x_2, x_1^2, x_1 x_2, x_2^2]^ \top$ are used for $\hat{\Sigma}_{\mathrm{vdp}}^{K_6}$, while $\hat{\Sigma}_{\mathrm{vdp}}^{K_{10}}$ employs a third-order extension of $g(x)$.
\begin{figure}[t]
    \vspace{-10pt}
    \includegraphics[width=1\columnwidth]{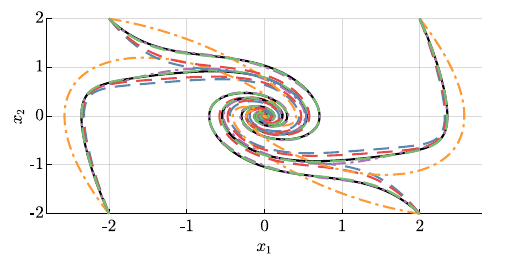}
    \vspace{-25pt}
    \caption{
    State responses of the CT simulation of $\Sigma_{\mathrm{vdp}}^{0.5}$, where (\crule{0, 0, 0}{6pt}{1pt}) is obtained with the true system, \mbox{(\crule{0.3467, 0.5360, 0.6907}{3pt}{1pt} \crule{0.3467, 0.5360, 0.6907}{3pt}{1pt})}, \mbox{(\crule{0.9153, 0.2816, 0.2878}{3pt}{1pt} \crule{0.9153, 0.2816, 0.2878}{3pt}{1pt})} and \mbox{(\crule{0.44157,   0.74902,   0.43216}{3pt}{1pt} \crule{0.44157,   0.74902,   0.43216}{3pt}{1pt})} with the respective OTFS surrogates $\hat{\Sigma}_{\mathrm{vdp}}^{\D^1}$, $\hat{\Sigma}_{\mathrm{vdp}}^{\D^2}$ and $\hat{\Sigma}_{\mathrm{vdp}}^{\D^3}$, and \mbox{(\crule{1, 0.5984, 0.2000}{3pt}{1pt} \crule{1, 0.5984, 0.2000}{1pt}{1pt} \crule{1, 0.5984, 0.2000}{3pt}{1pt})} with the data-driven Koopman model $\hat{\Sigma}_{\mathrm{vdp}}^{K}$.
    }\label{fig:surrogate05} \vspace{-5pt}
\end{figure}
We compare the surrogates with $\Sigma_{\mathrm{vdp}}^{0.5}$ by CT simulations under the same settings as in the previous experiment~\ref{sec:VDPexperiment1}. The results, displayed in Fig.~\ref{fig:surrogate05}, show that all the obtained surrogates approximate the behaviour of $\Sigma_{\mathrm{vdp}}^{0.5}$. Remarkably, $\hat{\Sigma}_{\mathrm{vdp}}^{\D^1}$ -trained with only 9 system snapshots- already outperforms the accuracy of the Koopman surrogate $\hat{\Sigma}_{\mathrm{vdp}}^{K}$. Furthermore, as predicted by Theorem~\ref{thm:universalApprox}, the accuracy of the OTFS models improves systematically with richer data dictionaries. Average RMSEs and runtimes are given in Tab.~\ref{tab:simulationMetrics}.
\begin{table}[t]
    \centering
    \caption{RMSE between the true system and the surrogate models and runtimes for the simulation results of Fig.~ \ref{fig:surrogate05}.}\label{tab:simulationMetrics}\vspace{-7pt}
    \begin{tabular}{lcc}
        Surrogate model                        & Average RMSE & Average runtime (s) \\
        \midrule
        $\hat{\Sigma}_{\mathrm{vdp}}^{\D^1}$   & 0.1187       & 0.0466              \\
        $\hat{\Sigma}_{\mathrm{vdp}}^{\D^2}$   & 0.0743       & 0.0477              \\
        $\hat{\Sigma}_{\mathrm{vdp}}^{\D^3}$   & 0.008        & 0.0493              \\
        $\hat{\Sigma}_{\mathrm{vdp}}^{K_6}$    & 0.2404       & 0.0054              \\
        $\hat{\Sigma}_{\mathrm{vdp}}^{K_{10}}$ & 0.0272       & 0.0062              \\
        \bottomrule
    \end{tabular}
\end{table}

\subsubsection{On-the-fly surrogation of the Van der Pol oscillator for a range of system parameter values}\label{sec:VDPexperiment3}
Third, we consider that the system parameter $\eta$ in~\eqref{eq:van_der_pol} lies within the range $\eta \in [0.3, 0.6]$. To capture the system dynamics for the different values of $\eta$, we define the observation set
\begin{equation}\label{eq:vdp_admissibleSet}
    \begin{aligned}
        \Theta_{\mathrm{vdp}}^4 = \{(x_1, x_2, \eta) \ \mid \ x_1, x_2 \in \mathbb{Z}, -2 \le x_1 \le 2, \\
        -2 \le x_2 \le 2, \; \eta \in \{0.3, 0.5, 0.6\}\},
    \end{aligned}
\end{equation}
and obtain the data dictionary $\D_{\mathrm{vdp}}^4$.
Now, we obtain the surrogate  $\hat{\Sigma}_{\mathrm{vdp}}^{\D^4}$ and compare in simulation with the true nonlinear system, using the same settings for the interpolant and the simulations described in the second experiment~\ref{sec:VDPexperiment2}, and CS38 with 6 equidistant intervals. In addition, for these simulations we fix $\eta$ to different values contained in~\eqref{eq:vdp_admissibleSet} that are not included in $\Theta_{\mathrm{vdp}}^2$. In this case, \texttt{ode45} requested a maximum of 978 evaluations of the interpolant, leading to an average runtime and RMSE of 0.048 seconds and 0.106, respectively.
\begin{figure}[b]
    \includegraphics[width=1\columnwidth]{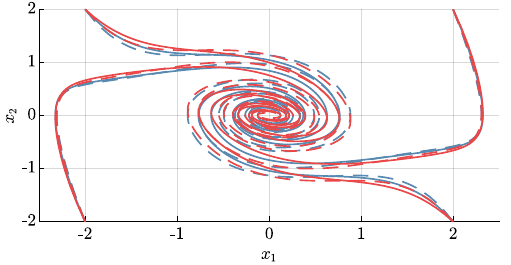}
    \vspace{-22pt}
    \caption{State responses of the CT simulation of the Van der Pol oscillator, where (\crule{0.3467, 0.5360, 0.6907}{6pt}{1pt}) and (\crule{0.9153, 0.2816, 0.2878}{6pt}{1pt}) are obtained with the true nonlinear systems ${\Sigma}_{\mathrm{vdp}}^{0.35}$ and ${\Sigma}_{\mathrm{vdp}}^{0.47}$, respectively, and (\crule{0.3467, 0.5360, 0.6907}{3pt}{1pt} \crule{0.3467, 0.5360, 0.6907}{3pt}{1pt}) and (\crule{0.9153, 0.2816, 0.2878}{3pt}{1pt} \crule{0.9153, 0.2816, 0.2878}{3pt}{1pt}) are obtained with the surrogate $\hat{\Sigma}_{\mathrm{vdp}}^{\D^2}$ evaluated at $\eta =0.35$ and $\eta = 0.47$, respectively.} \label{fig:surrogateVariable} \vspace{-17pt}
\end{figure}
These results are displayed in Fig.~\ref{fig:surrogateVariable}.

\vspace{-5pt}\subsection{Interconnection of MSD systems}
In this example,
we consider an interconnection of MSD systems
depicted in Fig.~\ref{fig:MSDsketch},
whose dynamics are given by:
\begin{figure}[t]
    \centering
    \scalebox{1}[0.95]{\includegraphics[width=0.85\columnwidth]{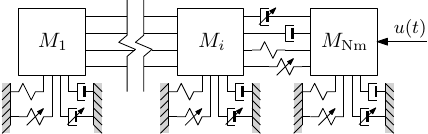}}
    \caption{Sketch of the interconnection of MSD systems.} \label{fig:MSDsketch}\vspace{-10pt}
\end{figure}
\begin{equation}
    \Sigma_{\mathrm{msd}} \coloneq \left\{
    \begin{aligned}
         & m \ddot{x}_1  = -F_1 - F_{1, 2},                                                          \\
         & m\ddot{x}_i   = -F_i - F_{i, \, i-1} - F_{i, \, i+1},                                     \\
         & m \ddot{x}_{n_\mathrm{m}}  = -F_{n_\mathrm{m}} - F_{n_\mathrm{m}, \, n_\mathrm{m}-1} + u,
    \end{aligned}
    \right.
\end{equation}
for $2 \leq i \leq n_\mathrm{m}-1$, where $F_{i, \, j}$ and $F_i$ are the forces applied to the $i$-th mass by the adjacent masses and from the connection to the rigid wall, respectively, given by
\begin{equation*}
    \begin{split}
        F_{i, j} & = k_1 (x_i \!-\! x_j)\!+\!b_1 (x_i\!-\!x_j)\!+\!(x_i\!-\!x_j)^3\!+\!b_2(\dot{x}_i\!-\!\dot{x}_j)^3, \\
        F_i      & = k_1 x_i + b_1 \dot{x}_i,
    \end{split}
\end{equation*}
and the system parameters are chosen as the number of masses $n_\mathrm{m}=5$, $m=1$ kg, $k_1=0.5$ N/m, $b_1=1$ Ns/m, $b_2=2$ Ns/m, and the external force $u$ applied to the last mass is in Newtons.
Further, we define the admissible space $\mathcal{Z}_{\mathrm{msd}} := \mathcal{X}_{\mathrm{msd}} \times \mathcal{U}_{\mathrm{msd}}$, where \mbox{$\mathcal{X}_{\mathrm{msd}} := \{\{-2.2 \leq x_i \leq  2.2 \}_{i=1}^5, \{-1.5 \leq x_i \leq  1.5 \}_{i=6}^{10} \}$}, \mbox{$\mathcal{U}_{\mathrm{msd}} = \{-1.5 \leq u  \leq  1.5 \}$},
and we construct the data set $\D_{\mathrm{msd}}$ with $N=100$ distinct system linearizations at randomly drawn points uniformly distributed within $\mathcal{Z}_{\mathrm{msd}}$.
In contrast, as the respective state-order and input dimension of $\Sigma_{\mathrm{msd}}$ are $\nx = 10$ and \mbox{$\dnu = 1$}, a regular grid containing only the vertices of the hyper-rectangle described by $\mathcal{Z}_{\mathrm{msd}}$ requires $2^{\nx+\dnu}\!=\!2048$ system snapshots.

Similar as in Subsection~\ref{sec:VDPexperiment2}, we construct the interpolant $\mathcal{I}_{\mathrm{msd}}^{\D_{\mathrm{msd}}}$, where $\D_{\mathrm{msd}}$, the RBF $\phi$ in~\eqref{eq:RBFhardy}, the polynomial basis $\mathcal{I}_{\mathrm{poly}}(x, u) = \beta_1 + \beta_2 x_1 + \dots \beta_{11} x_{10} + \beta_{12} u$, and the width parameter $c$ obtained by minimizing~\eqref{eq:optimization} with Bayesian optimization are the inputs to Algorithm~\ref{alg:interp}. Then, we assemble the surrogate $\hat{\Sigma}_{\mathrm{msd}}^{\D_{\mathrm{msd}}}$ according to Algorithm~\ref{alg:otfs} using $\mathcal{I}_{\mathrm{msd}}^{\D_{\mathrm{msd}}}$ and CS38 with 6 equidistant intervals.

\begin{figure}[t]
    \includegraphics[width=0.95\columnwidth]{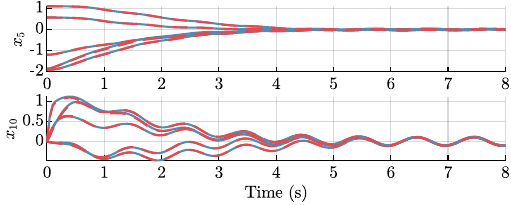}
    \vspace{-10pt}
    \caption{Simulated position (top) and velocity (bottom) trajectories of the $5$-th mass of the MSD interconnection, where (\crule{0.3467, 0.5360, 0.6907}{6pt}{1pt}) and (\crule{0.9153, 0.2816, 0.2878}{3pt}{1pt} \crule{0.9153, 0.2816, 0.2878}{3pt}{1pt}) are obtained from the nonlinear system $\Sigma_{\mathrm{msd}}$ and the surrogate $\hat{\Sigma}_{\mathrm{msd}}^{\D_{\mathrm{msd}}}$, respectively.}\label{fig:MSDoutput} \vspace{-5pt}
\end{figure}

Next, we test $\hat{\Sigma}_{\mathrm{msd}}^{\D_{\mathrm{msd}}}$ by executing one thousand CT simulations, where the states corresponding to the mass velocities, i.e., $\{x_i\}_{i=6}^{10}$, are initialized at zero, and the states corresponding to the mass positions, i.e., $\{x_i\}_{i=1}^5$, are initialized at random values drawn within $\mathcal{Z}_{\mathrm{msd}}$ with a uniform distribution. Moreover, the external force $u$ is set to $u(t) = 0.7 \sin(2\pi t)$, and each simulation is executed for 8 seconds using \texttt{ode45} with the default parameters and step size. For these simulations, \texttt{ode45} requested an average of 2352 evaluations of $\mathcal{I}_{\mathrm{msd}}^{\D_{\mathrm{msd}}}$, leading to an average computation time of 0.18 seconds. In Fig.~\ref{fig:MSDoutput}
\begin{figure}[t]
    \includegraphics[width=1\columnwidth]{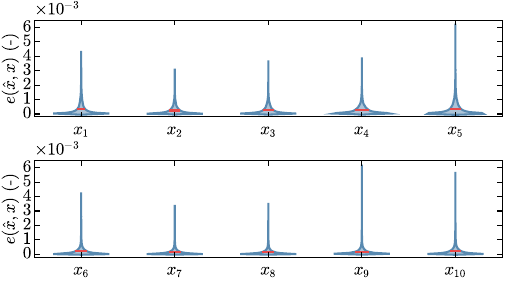}
    \vspace{-22pt}
    \caption{Estimated pdf of the RMSE over one thousand simulations of $\hat{\Sigma}_{\mathrm{msd}}^{\D_{\mathrm{msd}}}$ w.r.t. $\Sigma_{\mathrm{msd}}$, denoted as $e(\cdot, \cdot)$, where the pdf is computed based on a normal kernel function and (\crule{0.9153, 0.2816, 0.2878}{6pt}{1pt}) indicates the mean of the error.} \label{fig:MSDprobabilistic} \vspace{-5pt}
    \vspace{-10pt}
\end{figure}
we display the trajectory of the last mass obtained from some simulation results, where it is observed that the surrogate $\hat{\Sigma}_{\mathrm{msd}}^{\D_{\mathrm{msd}}}$ approximates true nonlinear system $\Sigma_{\mathrm{msd}}$ with remarkable accuracy. Furthermore, we compute the RMSE for each state of the simulated $\hat{\Sigma}_{\mathrm{msd}}^{\D_{\mathrm{msd}}}$ w.r.t. the simulated $\Sigma_{\mathrm{msd}}$. Then, we compute the estimated \emph{probability density function} (pdf) of the RMSE data based on a normal kernel function~\cite{botevKernelDensityEstimation2010}, and besides some outliers with error values of up to $6\cdot10^{-3}$, most of the error across all simulations is below $10^{-3}$, as shown in Fig.~\ref{fig:MSDprobabilistic}.
\vspace{-5pt}
\section{Conclusions\label{sec:conclusion}}\vspace{-2pt}
We have presented a novel method that, based on a finite set of system linearizations, which can be efficiently obtained from some simulation environments even if the explicit equations are not available, approximates the underlying nonlinear dynamics with a surrogate model that is
computed ``on-the-fly'' when required. Compared to other methods, the main advantages of our approach are: (i) it approximates the \emph{global} instead of the \emph{local} underlying nonlinear dynamics, and we show that the proposed surrogate model converges to the true nonlinear system as the number of system linearizations increases; (ii) it avoids computationally intensive processes, such as the training of neural networks, and (iii) the inclusion of new system observations only requires solving a linear least squares problem and is therefore straightforward.
In addition, we have demonstrated the global approximation capabilities of our method in various simulation experiments. For further research, we plan to investigate error estimates for the proposed surrogate modelling approach, the combination with state-order reduction techniques to enhance the scalability of our approach and to introduce sparsity in the interpolant $\mathcal{I}^\D$ to reduce the computational cost.
\vspace{-5pt}

\bibliographystyle{ieeetr}
\bibliography{on-the-fly-surrogation}

\providecommand{\noopsort}[1]{}
\begin{thebibliography}{10}

\bibitem{7823045}
T.~Samad, ``A survey on industry impact and challenges thereof,'' {\em IEEE Control Systems Magazine}, vol.~37, no.~1, pp.~17--18, 2017.

\bibitem{tothModelingIdentificationLinear2010}
R.~T{\'o}th, {\em Modeling and {{Identification}} of {{Linear Parameter-Varying Systems}}}, vol.~403 of {\em Lecture {{Notes}} in {{Control}} and {{Information Sciences}}}.
\newblock Springer, 2010.

\bibitem{whiteLinearParameterVaryingControl2013}
A.~P. White, G.~Zhu, and J.~Choi, {\em Linear {{Parameter-Varying Control}} for {{Engineering Applications}}}.
\newblock {{SpringerBriefs}} in {{Electrical}} and {{Computer Engineering}}, Springer, 2013.

\bibitem{doi:10.1137/16M1062296}
J.~L. Proctor, S.~L. Brunton, and J.~N. Kutz, ``Generalizing {Koopman} theory to allow for inputs and control,'' {\em SIAM Journal on Applied Dynamical Systems}, vol.~17, no.~1, pp.~909--930, 2018.

\bibitem{mauroyKoopmanOperatorSystems2020}
A.~Mauroy, I.~Mezi{\'c}, and Y.~Susuki, eds., {\em The {{Koopman Operator}} in {{Systems}} and {{Control}}: {{Concepts}}, {{Methodologies}}, and {{Applications}}}, vol.~484 of {\em Lecture {{Notes}} in {{Control}} and {{Information Sciences}}}.
\newblock Springer, 2020.

\bibitem{zhuPhysicsConstrainedDeepLearning2019}
Y.~Zhu, N.~Zabaras, P.-S. Koutsourelakis, and P.~Perdikaris, ``Physics-{{Constrained Deep Learning}} for {{High-dimensional Surrogate Modeling}} and {{Uncertainty Quantification}} without {{Labeled Data}},'' {\em Journal of Computational Physics}, vol.~394, pp.~56--81, 2019.

\bibitem{10383457}
M.~Aguiar, A.~Das, and K.~H. Johansson, ``Universal approximation of flows of control systems by recurrent neural networks,'' in {\em Proc. of the 62nd {{IEEE}} Conf. on Decision and Control}, pp.~2320--2327, 2023.

\bibitem{antonelloPhysicsInformedNeural2023}
F.~Antonello, J.~Buongiorno, and E.~Zio, ``Physics informed neural networks for surrogate modeling of accidental scenarios in nuclear power plants,'' {\em Nuclear Engineering and Technology}, vol.~55, no.~9, pp.~3409--3416, 2023.

\bibitem{maclaurin2015autograd}
D.~Maclaurin, D.~Duvenaud, and R.~P. Adams, ``Autograd: {{Effortless}} gradients in {Num}{Py},'' in {\em {{ICML}} {{AutoML}} Workshop}, vol.~238, 2015.

\bibitem{zhangLocalLTIModel2020}
Q.~Zhang, L.~Ljung, and R.~Pintelon, ``On {{Local LTI Model Coherence}} for {{LPV Interpolation}},'' {\em IEEE Transactions on Automatic Control}, vol.~65, pp.~3671--3676, Aug. 2020.

\bibitem{BACHNAS2014272}
A.~Bachnas, R.~T{\'o}th, J.~Ludlage, and A.~Mesbah, ``A review on data-driven linear parameter-varying modeling approaches: {{A}} high-purity distillation column case study,'' {\em Journal of Process Control}, vol.~24, no.~4, pp.~272--285, 2014.

\bibitem{apostolCalculus1Onevariable1980}
T.~M. Apostol, {\em Calculus. 1: {{One-variable}} Calculus, with an Introduction to Linear Algebra}.
\newblock Wiley, 1980.

\bibitem{wendlandScatteredDataApproximation2004}
H.~Wendland, {\em Scattered {{Data Approximation}}}.
\newblock Cambridge University Press, 1~ed., 2004.

\bibitem{fasshauerMeshfreeApproximationMethods2008}
G.~E. Fasshauer, {\em Meshfree Approximation Methods with {{MATLAB}}}.
\newblock No.~6 in Interdisciplinary Mathematical Sciences, New Jersey: World Scientific, 2008.

\bibitem{koelewijnAnalysisControlNonlinear2023}
P.~J.~W. Koelewijn, {\em Analysis and Control of Nonlinear Systems with Stability and Performance Guarantees: {{A}} Linear Parameter-Varying Approach}.
\newblock PhD thesis, Eindhoven University of Technology, 2023.

\bibitem{oluchaAutomatedLinearParameterVarying2025}
E.~J. Olucha, P.~J.~W. Koelewijn, A.~Das, and R.~T{\'o}th, ``Automated {{Linear Parameter-Varying Modeling}} of {{Nonlinear Systems}}: {{A Global Embedding Approach}},'' in {\em Proc. of the 6th IFAC LPVS Workshop}, 2025.

\bibitem{Tibshirani:1996fxl}
R.~Tibshirani, ``Regression shrinkage and selection via the {Lasso},'' {\em J. Roy. Statist. Soc. B}, vol.~58, no.~1, pp.~267--288, 1996.

\bibitem{specht1991general}
D.~F. Specht {\em et~al.}, ``A general regression neural network,'' {\em IEEE Transactions on neural networks}, vol.~2, no.~6, pp.~568--576, 1991.

\bibitem{wu_local_1993}
Z.-M. Wu and R.~Schaback, ``Local error estimates for radial basis function interpolation of scattered data,'' {\em IMA Journal of Numerical Analysis}, vol.~13, no.~1, pp.~13--27, 1993.

\bibitem{amidrorScatteredDataInterpolation2002}
I.~Amidror, ``Scattered data interpolation methods for electronic imaging systems: A survey,'' {\em Journal of Electronic Imaging}, vol.~2, no.~2, pp.~157--176, 2002.

\bibitem{BJORCK1990465}
{\AA}.~Bj{\"o}rck, ``Least squares methods,'' vol.~1 of {\em Handbook of Numerical Analysis}, pp.~465--652, Elsevier, 1990.

\bibitem{nevistić_primbs_1996}
V.~Nevisti{\'c} and J.~A. Primbs, ``Constrained nonlinear optimal control: A converse {{HJB}} approach,'' {\em California Inst. of Technol., Tech. Rep.}, Dec. 1996.

\bibitem{atkinsonIntroductionNumericalAnalysis1989}
K.~E. Atkinson, {\em An Introduction to Numerical Analysis}.
\newblock Wiley, 2~ed., 1989.

\bibitem{brochuTutorialBayesianOptimization2010}
E.~Brochu, V.~M. Cora, and N.~de~Freitas, ``A tutorial on {Bayesian} optimization of expensive cost functions, with application to active user modeling and hierarchical reinforcement learning.'' arXiv: 1012.2599, 2010.

\bibitem{williamsDataDrivenApproximation2015}
M.~O. Williams, I.~G. Kevrekidis, and C.~W. Rowley, ``A {{Data}}--{{Driven Approximation}} of the {{Koopman Operator}}: {{Extending Dynamic Mode Decomposition}},'' {\em Journal of Nonlinear Science}, vol.~25, pp.~1307--1346, Dec. 2015.

\bibitem{Pan2024}
S.~Pan, E.~Kaiser, B.~M. de~Silva, J.~N. Kutz, and S.~L. Brunton, ``{Py}{Koopman}: A {Python} {Package} for {Data}-{Driven} {Approximation} of the {Koopman} {Operator},'' {\em Journal of Open Source Software}, vol.~9, no.~94, p.~5881, 2024.

\bibitem{botevKernelDensityEstimation2010}
Z.~I. Botev, J.~F. Grotowski, and D.~P. Kroese, ``Kernel density estimation via diffusion,'' {\em The Annals of Statistics}, vol.~38, no.~5, pp.~2916--2957, 2010.

\end{thebibliography}


\addtolength{\textheight}{-12cm}

\end{document}